\RequirePackage{lineno} 

\documentclass{article}
\usepackage{fancyhdr}

\usepackage[round]{natbib}
\usepackage{graphicx}
\usepackage{multirow}
\usepackage{color}
\usepackage{amssymb}
\usepackage{amsmath}
\usepackage{enumerate}
\usepackage{lineno}
\usepackage{caption}
\usepackage{textcomp}
\usepackage{gensymb}
\usepackage{upgreek}
\usepackage{float}
\usepackage{lipsum}
\usepackage{epstopdf, epsfig}
\usepackage{authblk}
\usepackage{rotating}

\usepackage[letterpaper, margin=1.0 in]{geometry}

\begin{document}
	
	\title{Experimental evidence for a universal threshold characterizing wave-induced sea ice break-up}
	\author[1]{J.J. Voermans}
	\author[2,3]{J. Rabault}
	\author[4]{K. Filchuk}
	\author[4]{I. Ryzhov}
	\author[5]{P. Heil}
	\author[6]{A. Marchenko}
	\author[7]{C. Collins}
	\author[8]{M. Dabboor}
	\author[8]{G. Sutherland}
	\author[1,9]{A.V. Babanin}
	\affil[1]{University of Melbourne, Australia}
	\affil[2]{Norwegian Meteorological Institute, Norway}
	\affil[3]{University of Oslo, Norway}
	\affil[4]{Arctic and Antarctic Research Institute, Russia}
	\affil[5]{Australian Antarctic Division and Australian Antarctic Program Partnership, University of Tasmania, Australia}
	\affil[6]{The University Centre in Svalbard, Norway}
	\affil[7]{U.S. Army Engineering Research and Development Center, USA}
	\affil[8]{Environment and Climate Change Canada, Canada}
	\affil[9]{Laboratory for Regional Oceanography and Numerical Modeling, National Laboratory for Marine Science and Technology, Qingdao, China}
	\setcounter{Maxaffil}{0}
	\renewcommand\Affilfont{\itshape\small}
	\date{}
	\maketitle

	\begin{abstract}
		Waves can drastically transform a sea ice cover by inducing break-up over vast distances in the course of a few hours. However, relatively few detailed studies have described this phenomenon in a quantitative manner, and the process of sea ice break-up by waves needs to be further parameterized and verified before it can be reliably included in forecasting models. In the present work, we discuss sea ice break-up parameterization and demonstrate the existence of an observational threshold separating breaking and non-breaking cases. This threshold is based on information from two recent field campaigns, supplemented with existing observations of sea ice break-up. The data used cover a wide range of scales, from laboratory-grown sea ice to polar field observations. Remarkably, we show that both field and laboratory observations tend to converge to a single quantitative threshold at which the wave-induced sea ice break-up takes place, which opens a promising avenue for robust parametrization in operational forecasting models.
	\end{abstract}
	
	\section{Introduction}
	Surface gravity waves can propagate tens to hundreds of kilometers into the ice pack before the ice fully dissipates their energy \citep[e.g.,][]{Kohout14,Stopa18}. In the process, waves flex the ice, imposing stresses on the elastic and brittle ice sheet. When these stresses exceed a critical value, the sea ice will crack or break, creating large regions of broken ice floes with complex dynamics \citep{horvat2016interaction, hwang2017winter}. Once broken, the ice is able to move more freely, reducing the attenuation of wave energy \citep[e.g.,][]{Collins15}, and thereby allowing waves to penetrate even further into the ice pack. This drives a series of secondary processes in the coupled air-sea system that can further affect the properties of the ice, including enhanced upper ocean mixing in sea ice covered waters \citep{Thomas19}, sea ice drift \citep{Boutin20}, and lateral melting of ice floes \citep{Steele92}. Hence, the extent to which waves can impact the morphology of the sea ice cover is defined by the balance between wave energy dissipation as a function of sea ice properties on the one hand, and the break-up of the sea ice by the stresses imposed onto the ice by the waves on the other hand. Evidently, the complex and coupled processes of ice-induced wave attenuation and wave-induced sea ice breakup need to be understood, quantified, and modeled, before wave-ice interaction processes can be reasonably implemented in operational forecasting models.
	
	Studies have, so far, mainly focused on the attenuation of wave energy in sea ice covers and identified a series of conservative and dissipative processes that damp wave energy in sea ice. These include wave scattering \citep[e.g.,][]{vaughan2007scattering, meylan2018three}, stresses within the ice layer \citep[e.g.,][]{Wang10,Sutherland19_twolayer}, turbulence \citep{Liu88,Voermans19}, brine migration \citep{Marchenko17}, and interactions between ice-floes \citep{Rabault19,Herman19}. Although there is still debate regarding when and where these processes are important \citep{Thomson18,Squire20}, they have been, to various degrees, parameterized, validated, and/or implemented in numerical wave models \citep[e.g.,][]{WWIII}. Our understanding of wave-induced sea ice breakup is, however, significantly lacking, and few studies are available \citep[with the notable exception of the studies by][]{Crocker89,Langhorne98,Dumont11,Williams13a}.
	
	Fundamentally, wave-induced sea ice break-up is determined by a large set of highly environmental dependent wave and ice parameters. Those include the mechanical properties of sea ice (the flexural strength of the ice $\sigma$, elastic or Young's modulus $Y$), its material properties (ice salinity $S_{ice}$, ice temperature $T_i$, water $\rho_w$ and ice density $\rho_{ice}$), the scale of the ice (ice thickness $h$ and horizontal length scale of the ice $L_{ice}$), as well as wave field characteristics (wave amplitude $a$ and wave length $\lambda$), the gravitational acceleration $g$ and time $t$. We ignore surface tension and viscosity here due to the large length scales associated with the problem, though it is acknowledged that the ice viscosity could potentially play a role. We also ignore $L_{ice}$, the floe size, and focus on solid ice instead, that is, $L_{ice} \gg \lambda$. If we also consider the ice to be flexible enough to follow the wave surface reasonably well, that is, the ice is not thick enough to be rigid at the length scale of the wavelength, buoyancy effects might be ignored such that $\rho_w$, $\rho_{ice}$, and $g$, are only of minor importance. The ice mechanical properties $\sigma$ and $Y$ are, perhaps, the most complex variables in this set as they are strongly related to the environmental conditions to which the was exposed at its formation and during the rest of its lifetime. In particular, exposure to the cyclic bending of the ice by waves can lower the flexural strength of the ice \citep[e.g.,][]{Langhorne98}, commonly known as fatigue, but can also strengthen the ice when steady stress loads are applied to the ice \citep{Murdza20}, such as by wind and currents, whereas local heterogeneities in sea ice can lead to localized concentration of stresses. While these complexities are intrinsic to the physics of the wave-induced sea ice break-up problem, a full understanding of these processes are outside the scope of this study. Here, we ignore the dependence of sea ice material properties with its history (or time $t$), and adopt the traditional dependence of $\sigma$ and $Y$ on the brine volume fraction of the ice $\upsilon_b$, which has been related to the temperature and salinity of the ice, such that $\sigma=f\left(S_{ice},T_{ice}\right)$ and $Y=f\left(S_{ice},T_{ice}\right)$.
	
	If we then define the wave-induced sea ice break-up similitude by a non-dimensional parameter $I_{br}$ using the Pi-theorem, the break-up problem can be formulated as:
	
	\begin{equation}
	I_{br}=f\left( \frac{\sigma}{Y},\frac{a}{\lambda},\frac{h}{\lambda} \right).
	\label{eq:Dimensional}
	\end{equation}
	
	\noindent where $\sigma/Y$ is the strain, $a/\lambda$ is the wave steepness and $h/\lambda$ is the relative ice thickness. The dependency of $I_{br}$ on these parameters can be determined by considering the ice sheet as an elastic plate. This results in the flexural strain
	
	\begin{equation}
	\varepsilon=\frac{h}{2}\frac{\partial^2\eta}{\partial x^2},
	\label{eq:strain_material}
	\end{equation}
	
	\noindent where $\eta$ is the wave surface elevation in the horizontal direction $x$. Considering a periodic wave $\eta=a\sin(kx-\omega t)$, where $k=2\pi/\lambda$ is the wave number and $\omega$ is the radian wave frequency, the maximum strain is defined as \citep[e.g.][]{Dumont11}:
	
	\begin{equation}
	\varepsilon=\frac{2\pi^2 a h}{\lambda^2}.
	\label{eq:strai_wave}
	\end{equation}
	
	Assuming elastic behaviour of the ice layer, the strain can be considered proportional to the flexural strength $\sigma$ of the ice, leading to $\varepsilon=\sigma/Y$. It then follows that a monochromatic wave will break the ice when $2\pi^2ahY/\sigma\lambda^2>1$. The wave-induced sea ice break-up parameter $I_{br}$ is, therefore:
	
	\begin{equation}
	I_{br}=\frac{ahY}{\sigma\lambda^2}.
	\label{eq:breakup_number}
	\end{equation}
	
	This break-up parameter is consistent with Eqn.\hspace{1mm}(\ref{eq:Dimensional}), and forms the basis of the recent wave-induced sea ice break-up scheme implemented in coupled wave-ice models \citep{Dumont11,Williams13a,Williams13b,Ardhuin18,Boutin18,Boutin20}. It follows from Eqn.\hspace{1mm}(\ref{eq:strai_wave}) and Eqn.\hspace{1mm}(\ref{eq:breakup_number}) that the break-up threshold for a monochromatic wave is approximately $I_{br}=1/2\pi^2 \approx 0.05$, or, strictly speaking, when fatigue and local sea ice heterogeneities are considered $I_{br}\leq 0.05$. \citet{Boutin18} proposed a threshold 3.6 times smaller, i.e. $I_{br}=0.014$, based on statistical considerations that the relative maximum strain of a Gaussian random sea state is larger than that of a monochromatic wave. However, to the best of our knowledge, no study has extensively validated the value of the critical threshold $I_{br}$, nor its universality across a wide range of wave and ice scales. Without convincing validation, the value of this threshold remains an ambiguous extra degree of freedom needed to configure the model and to fit to observations, making it difficult to confidently apply the model at a global scale.
	
	Currently, the lack of a large number of wave-induced sea ice break-up observations, and the uncertainties associated with these, are arguably the foremost reasons for the uncertainty in parameterizing wave-induced sea ice break-up. Measuring wave and ice properties in the harsh polar environment is challenging, both logistically and technically, even in perfect weather conditions -- itself a rare event -- especially considering that sea ice break-up often happens during storms. Observing sea ice break-up requires either continuous visual observations, or refined experimental techniques. Even in the event that sea ice break-up is observed, identification of the exact instant at which the ice breaks (that is, the individual wave responsible for the break-up event) is problematic, as it does not necessarily identify the critical threshold of $I_{br}$, but rather presents a sufficient condition for break-up. That means, that if a wave with known amplitude is observed in the sea ice cover and triggers ice break-up, all what is known is that any wave with the same wave length and an amplitude equal to or larger than the amplitude recorded will break the ice. The contrapositive is true for any wave-induced ice motion taking place without breaking the ice cover. This is further complicated by the deterministic nature of the break-up event itself, that is, in theory we could measure the exact wave event responsible for the break-up, while, in contrast, the identified wave event is a result of the incoherent nature of the wave field and is, therefore, related to the statistical properties of the wave field instead. To bring light on this question, we suggest that many observations of wave-induced sea ice break-up and wave-induced sea ice motion without break-up should be collected. Then, if there should exist a critical, universal threshold for $I_{br}$ as defined in Eqn.\hspace{1mm}(\ref{eq:breakup_number}), a clear separation between unbroken and broken ice conditions should be observed, independently of the details of the ice conditions.
	
	In this study, we attempt to perform such an analysis. For this, we use the results of wave-induced ice motion measurements from two recent field campaigns, one in the Antarctic and the other in the Arctic. In addition, the data obtained are also complemented with an extensive set of observations from both laboratory and field experiments, collected throughout the literature. Thereafter, we approximate the critical wave-induced sea ice break-up criterion based on all data combined, and identify a universal threshold for $I_{br}$. 
	
	\section{Methods}
	\label{sec:Methods}
	
	\subsection{Field Experiments}
	
	\begin{figure}
		\centerline{\includegraphics[]{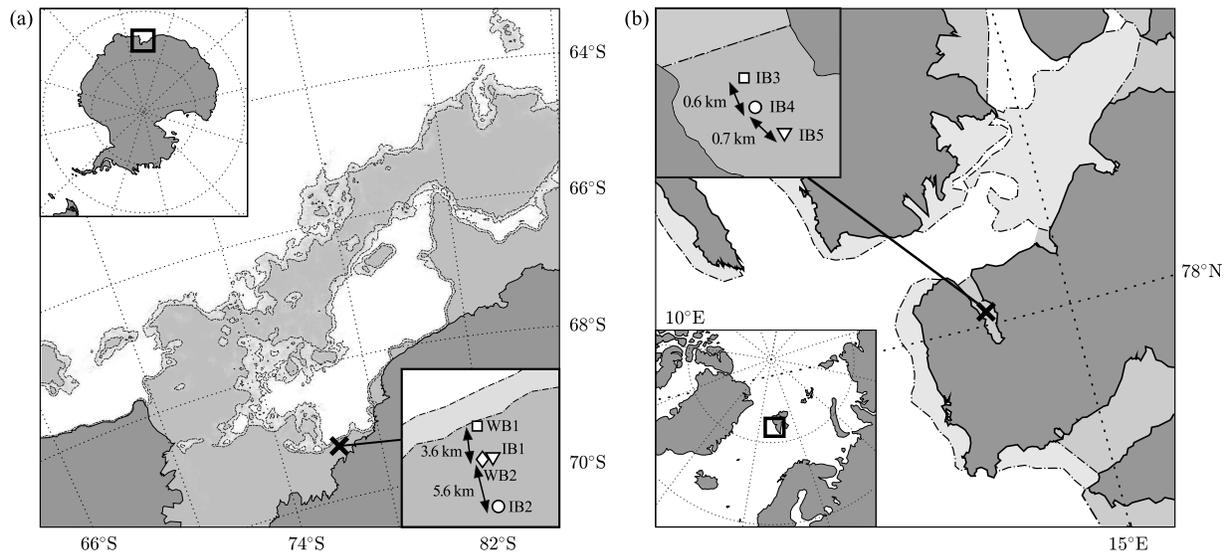}}
		\caption{Map of the field experiment sites on (a) Antarctic fast sea ice and (b) fast ice in Svalbard. Deployment sites are indicated by a cross. Continents are shaded dark gray, whereas sea ice concentration is represented by the light gray shades using two contour levels, indicative of (a) sea ice concentration of 25\% and 75\% derived from AMSR2 for 02-Jan-2020 \citep{Spreen08}, for light and dark grey respectively, and (b) open drift ice and very close drift ice obtained from the Norwegian Meteorological Institute Ice Service for 23-Mar-2020, for light and dark grey respectively. Instruments were deployed along a line perpendicular to the unbroken ice edge (see insets), and consisted of wave buoys (WB: Spotter buoys, Sofar Ocean Technologies) and open source ice motion loggers \citep[referred to as ice buoys, IB;][]{Rabault20}. Note that in (a), IB1 is shifted laterally for visualization purposes but in reality it is only 40 m apart from WB2.}
		\label{fig:Map_Antarctica}
	\end{figure}
	
	In the present study, the focus is on data from two recent field experiments, aiming to measure the wave-induced ice motion which lead to sea ice break-up. The first experiment took place in the Antarctic ice pack, and the second in the Arctic ice pack.
	
	\subsubsection{Deployment in the Antarctic}
	
	The Antarctic deployments occurred on (land)fast ice on the eastern rim of the Amery Ice Shelf (69.3$^\circ$S 76.3$^\circ$E, see Fig. \ref{fig:Map_Antarctica}a) on the 7th of December 2019. The instruments deployed consisted of two wave buoys, denoted as WB in the following (Spotter buoys, from Sofar Ocean Technologies), and two low-cost open-source ice motion loggers \citep[][hereafter referred to as ice buoys and denoted IB]{Rabault20}. Both the wave and ice buoys are compact solar-charged position and motion recording instruments with real-time Iridium transmission capability. The wave buoys measure displacement at 2.5 Hz using GPS and transmit wave and position data at a user defined interval. For the deployment period considered here, only integral wave parameters and battery power status were transmitted every half an hour. The ice buoys measure the ice motion using an inertial motion unit (IMU) performing measurements at 10 Hz and transmit the full wave spectrum, geographical location and battery power status at a predefined interval, here, every 3 hours. The accuracy of the vertical displacement is approximately 0.02 m for the wave buoy. For high frequency waves, the accuracy of the ice buoy is $O$(mm) \citep{Rabault16}, but the noise level increases with decreasing wave frequencies \citep{Rabault20}. For more technical details on the wave and ice buoys the reader is referred to \citet{Raghukumar19} and \citet{Rabault20}, respectively.
	
	The instruments were deployed along a line perpendicular to the unbroken ice edge. The first wave buoy (WB1) is about 100-200 m from the edge (see inset Fig. \ref{fig:Map_Antarctica}a). The second wave buoy (WB2) and first ice buoy (IB1) are deployed 3.7 km from the solid ice edge, close to each other (the initial distance between WB2 and IB1 is around $40$ m), whereas the last ice buoy (IB2) was deployed about 9.3 km from the edge. Wave buoys were deployed closest to the solid ice edge as these buoys are capable of surviving in the open water. While the ice buoys have sufficient buoyancy to float, they are expected to malfunction quickly after entering the water. At the time of deployment, the ice was estimated to be between 1 and 1.2 m thick.
	
	No drift nor significant wave events were recorded for the first three weeks after deployment. On the 2nd of January 2020 the uniform fast ice, on which all instruments rested, broke, and all instruments drifted with the sea ice. In the weeks that followed, geographical location and vertical ice motion under the influence of waves were obtained until instruments stopped transmitting. End of transmission happened for IB2 on 22--Jan, for WB1 and WB2 on 1-Feb, and for IB1 on 10--Mar. It is noteworthy that WB2 reconnected on the 3rd of March for half a day. The wave buoys failed due to depleted batteries, most likely caused by snow or ice coverage of the solar panels. Considering that batteries of the ice buoys were still close to fully charged during the last transmissions received from both instruments, we suspect the ice buoys were damaged by the ice or ended being submerged under water between floes. As our interest is in wave-induced sea ice break-up, this study focuses on observations obtained from January 2 -- 8, which is the period over which initial sea ice break-up was observed for an extensive stretch of fast ice.
	
\begin{figure}
		\centerline{\includegraphics[]{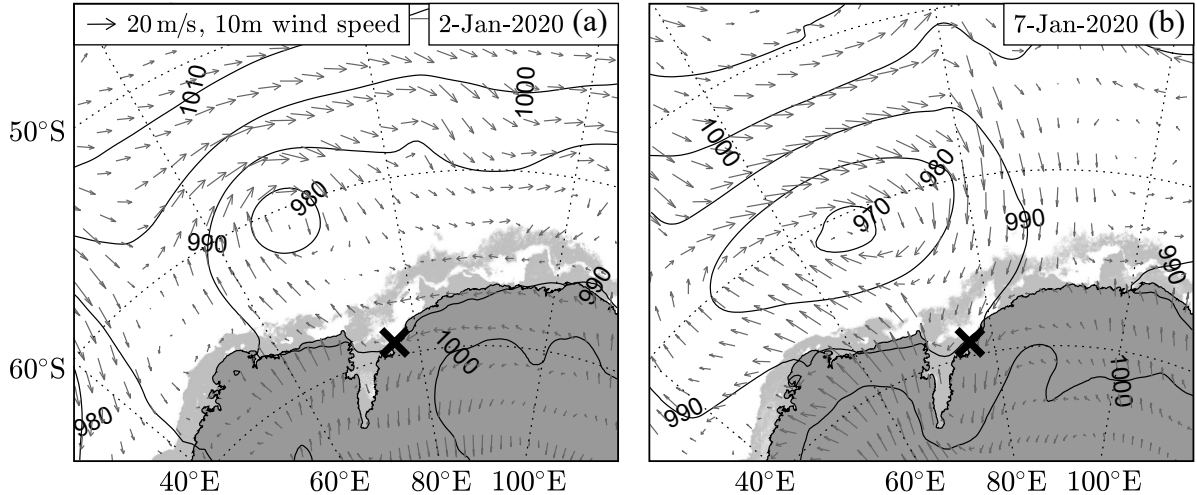}}
		\caption{Passage of two storms offshore of the Antarctic instrument site on (a) 2-Jan-2020 and (b) 7-Jan-2020. Contours show mean sea level pressure, and wind speed at 10 m is displayed through the vector field (both from ERA5). Light gray shading represents sea ice, derived from AMSR2 data \citep{Spreen08}. The black cross identifies the deployment site. The presence of relatively high wind speeds over the polynya region on 07-Jan-2020 is expected to generate wind waves at the deployment site.}
		\label{fig:Weather}
	\end{figure}

	During the first week of January, sea ice concentration is well represented by that shown in Fig. \ref{fig:Map_Antarctica}a. A polynya of approximately $100 km \times 300km$ separated the fast ice from a 100 km wide band of pack ice. Based on ERA5 re-analysis, three significant low pressure systems passed along the Antarctic continent over the time interval considered. The first merely skimmed the deployment site on the 2nd of January (Fig. \ref{fig:Weather}a), while the second moved north-east just before reaching the longitude of the instruments around January 5th. The third low-pressure system is expected to have the largest impact on the conditions near the deployment site, with an estimated wind speed of about 10 -- 15 m/s on the 7th of January (Fig. \ref{fig:Weather}b).
	
	\subsubsection{Deployment in the Arctic}
	
	The second field experiment was performed in Gr\o{}nfjorden, Svalbard (Fig. \ref{fig:Map_Antarctica}b). Three ice buoys were deployed on landfast sea ice between 10 and 13 March 2020, and recovered on the 28th of March. The unbroken ice edge was reasonably stable during the deployment, located roughly halfway through the fjord. The first ice buoy (IB3) was deployed approximately 500 m from the unbroken ice edge. The second (IB4) and third ice buoy (IB5) were deployed 600 m and 700 m apart. Ice thicknesses of 0.3 -- 0.4 m were measured along the main axis of the fjord at the start of the experiment. Based on the water temperature measured just under the ice and on the air temperature, the ice temperature is estimated to be about $-8^\circ$C. The salinity of the ice was determined by measuring the conductivity of melted sections of a 0.4 m long ice core, with bulk salinity of 0.68\%. Based on visual observations, the ice did not break during this field experiment.
	
	\subsection{Observations of sea ice break-up in previous literature}
	
	In addition to the ice motion and break-up observations collected during our field campaigns, a set of wave-induced break-up data was collected from the literature. Published data were used only when sufficient details about the wave and ice conditions were presented to determine $I_{br}$. Due to the near absence of concurrent measurements of all wave and ice properties, we consider it to be sufficient when ice thickness, wave height, and wave length (or wave period), are provided. The most critical requirement was that the published sea ice break-up event was, with sufficient confidence, attributed to the observed wave event. We exclude fresh water ice experiments and numerical studies. 
	
	The full dataset consists of 31 observations, including 14 wave events that did not result in ice break-up (9 of them from the laboratory), and 17 events where waves were responsible for the break-up of the ice (7 of which are laboratory observations). Besides the laboratory study of \citet{Herman18}, field observations were taken from \citet{Liu88}, \citet{Cathles09}, \citet{Marchenko11}, \citet{Marchenko12}, \citet{Asplin12}, \citet{Collins15}, \citet{Sutherland16}, \citet{Kohout16}, \citet{Marchenko19} and \citet{Kovalev20}. We note that the break-up observations made by \citet{Liu88} and \citet{Kohout16} are visual shipborne observations and not {\it in situ} measurements (see the complete set in Table \ref{tab:Table1}).
	
	In the case of the field experiment of \citet{Kovalev20}, wave conditions resulting in the largest $I_{br}$ were used here as these are the waves most likely responsible for the break-up event observed. For the field observations of \citet{Sutherland16}, cracks in the ice were argued to be responsible for the sudden change in the dispersion relation from flexural-gravity waves to gravity waves, and this transition is used here to determine the instant at which the ice was broken by waves. Additionally, the study of \citet{Cathles09} is included, and describes the impact of swell on the flexure of the Antarctic ice shelf. \citet{Cathles09} argue about the potential of most energetic swell events to promote crack propagation of the Nascent Iceberg. In a later study, \citet{Massom18} showed that there exists a strong correlation between the arrival of swell and the disintegration of the ice shelves. The ice motion amplitudes observed in \citet{Cathles09} are similar to those measured by \citet{Bromirski10}. While the ice shelf cannot be regarded as a thin ice sheet (and hence the validity of Eqn.\hspace{1mm}(\ref{eq:strain_material}) for this event can be questioned), this observation is, nevertheless, included for comparison reasons.
	
	As not all parameters were consistently and/or accurately measured across these studies, the uncertainty of the individual variables were estimated to approximate the uncertainty in $I_{br}$. Each variable was described by a triangular probability distribution, the most likely value of which is typically the value given in the respective study or, alternatively, the mean of the provided range. To obtain an uncertainty for the wave-induced sea ice break-up parameter, a large number of random values for each variable were generated and the 5th and 95th percentiles of $I_{br}$ were determined.
	
	For the wave amplitude, the most likely value is taken either as the cited wave amplitude, or half the significant wave height measured (i.e. $a=H_s/2$). For the wave period, if no specific period is provided, the (local) peak period is taken. For all direct observations of wave amplitude and wave period an uncertainty of 10\% is taken into account as the outer value of the triangular distribution, while for visual observations we use a larger uncertainty (case specific and dependent on the absolute values of the variables). Based on the water depth, either estimated or provided, the wave length is calculated following the linear dispersion relation. The impact of the ice on the wave length (i.e., the flexural, compressive, and ice added mass terms in the dispersion relation as expressed by for example \citet[][]{Sutherland16}) is assumed to be minor compared to the uncertainty included in the wave period. This is a reasonable assumption as most measurements have a wave period large than 7 s \citep[e.g.,][]{Sutherland16,Collins18}. As measurements of the ice thickness are expected to have higher uncertainty than the wave properties, an uncertainty of up to 50\% is considered, but larger values are chosen for shipborne visual observations.
	
	The mechanical properties of the ice have the largest uncertainty of all variables involved, in large part, as they are difficult to measure, particularly in this extreme environment. Only in the studies of \citet{Marchenko11,Marchenko12,Marchenko19} the flexural strength ($\sigma$) and/or Young's Modulus ($Y$) were measured {\it in situ} and therefore provide the narrowest range of uncertainty. Note that in the case of the tsunami wave observations of \citet{Marchenko12}, details of the ice properties during this experiment are provided in \citet{Marchenko13} and \citet{Karulina19}. For the Arctic field experiment (this study) and the observation of \citet{Asplin12} only ice salinity and temperature were measured. For these experiments we approximate $\sigma$ and $Y$ through their strong dependence on brine volume. Using the empirical relation of \citet{Frankenstein67}, the brine volume can be approximated by:
	
	\begin{equation}
	v_b=S_{ice}\left( \frac{49.185}{\left|T_{ice}\right|}+0.532\right),
	\label{eq:brine}
	\end{equation}
	
	\noindent where $v_b$ is the brine volume in ppt, $S_{ice}$ is the ice salinity in ppt, and $T_{ice}$ is the ice temperature in $^\circ$C. This gives an estimated sea ice brine volume of 4.51\% and 6.66\% during our Arctic experiment and the study of \citet{Asplin12}, respectively. As sea ice properties are strongly influenced by the conditions of its formation and development, the empirical relations for sea ice properties in terms of brine volume are considered to be region-specific \citep{Karulina19}. Hence, for our Arctic field experiment we consider empirical relations from the study of \citet{Karulina19}, which is focused on the ice properties in the Svalbard archipelago, yielding:
	
	\begin{equation}
	\sigma=0.5266 \exp\left(-2.804\sqrt{v_b}\right),
	\label{eq:sigma_Karulina}
	\end{equation}
	
	\begin{equation}
	Y=3.1031 \exp\left(-3.385\sqrt{v_b}\right),
	\label{eq:Y_Karulina}
	\end{equation}
	
	\noindent where the brine volume is in volume fraction instead of ppt here. The scatter of data for $\sigma$ and $Y$ it is in \citet{Karulina19} is used to quantify the uncertainty. For the sea ice break-up observation of \citet{Asplin12} we use the commonly used empirical relation of \citet{Timco94} instead to approximate the flexural strength:
	
	\begin{equation}
	\sigma=1.76 \exp\left(-5.88\sqrt{v_b}\right).
	\label{eq:sigma_Timco}
	\end{equation}
	
	For the Young's Modulus we consider the empirical relation of \citet{Vaudrey77}:
	
	\begin{equation}
	Y=5.31-0.436\sqrt{\upsilon_b}.
	\label{eq:Y_Vaudrey}
	\end{equation}
	
	Note that the unit of brine volume in Eqn.\hspace{1mm}(\ref{eq:sigma_Timco}) is in volume fraction whereas in Eqn.\hspace{1mm}(\ref{eq:Y_Vaudrey}) in ppt. It is worth mentioning that the value for $\sigma$ calculated following this approach in \citet{Asplin12} is incorrect due to a typographical error in their equation (compare Eqn.\hspace{1mm}(\ref{eq:sigma_Timco}) here to their Eqn.\hspace{1mm}(4)). An uncertainty of 50\% is assigned to $\sigma$ and $Y$ for the observation of \citet{Asplin12}.
	
	For all other observations where no details of sea ice properties were measured or provided, we assign a relatively conservative range of uncertainty to $\sigma$ and $Y$. For experiments within the Svalbard archipelago, we choose a range of $\sigma \in [0.109, 0.415]$ MPa and $Y \in [0.4, 3]$ GPa with most probable values of $\sigma=2.62$ MPa and $Y=1.25$ GPa \citep{Karulina19}. A wider range for $\sigma$ and $Y$ is expected to be found elsewhere and, as such, we expand the uncertainty for observations made in other regions given by $\sigma \in [0.1, 0.7]$ MPa and $Y \in [1, 6]$ GPa with most probable values of $\sigma=0.4$ MPa and $Y=3$. A summary of all data used and their estimated uncertainty is provided in Table \ref{tab:Table1}.
	
	\begin{sidewaystable}[]
		\begin{tabular}{llccccccc}
			\hline
			& ice status & $H_s$ (m)& $T$ (s) & $\lambda$ (m) & $d$ (m)& $h$ (m) & $Y$ (GPa) & $\sigma$ (MPa) \\
			\hline
			this study, Antarctica & break-up & $0.04\pm0.01$ & $5\pm0.5$ & $(32,39,47)$ & 450 & $(0.5,1,1.2)$ & $(1,3,6)$ & $0.4\pm0.3$ \\
			this study, Antarctica & no break-up & $0.05\pm0.005$ & $18.5\pm1.5$ & $(451,534,624)$ & 450 & $(0.5,1,1.2)$ & $(1,3,6)$ & $0.4\pm0.3$ \\
			this study, Arctic & no break-up & $0.1\pm0.01$ & $11.7\pm1.2$ & $(173,214,258)$ & 138 & $0.35\pm0.12$ & $(1,1.5,3)$ & $0.29\pm0.11$ \\
			this study, Arctic & no break-up & $0.04\pm0.004$ & $7.8\pm0.8$ & $(77,95,115)$ & 138 & $0.35\pm0.12$ & $(1,1.5,3)$ & $0.29\pm0.11$ \\
			this study, Arctic & no break-up & $0.07\pm0.007$ & $14.3\pm1.4$ & $(258,317,378)$ & 138 & $0.35\pm0.12$ & $(1,1.5,3)$ & $0.29\pm0.11$ \\
			\citet{Kovalev20} & break-up & $0.16\pm0.016$ & $8\pm0.8$ & $82\pm11$ & 15.3 & $0.45\pm0.15$ & $(1,3,6)$ & $0.2\pm0.07$ \\
			\citet{Asplin12} & break-up & $0.8\pm0.08$ & $13.5\pm1.4$ & $(230,285,344)$ & 1000 & $2\pm0.67$ & $(0.88,1.75,2.63)$ & $0.39\pm0.2^{(3)}$ \\
			\citet{Kohout16} & break-up & $(0.25,0.5,1)$ & $15\pm3$ & $(225,351,506)$ & 1000 & $(0.375,0.75,1.5)$ &	$(1,3,6)$ & $0.4\pm0.3$ \\
			\citet{Kohout16} & break-up & $(0.1,0.2,0.4)$ & $15\pm3$ & $(225,351,506)$ & 1000 & $(0.375,0.75,1.5)$ & $(1,3,6)$ & $0.4\pm0.3$ \\
			\citet{Cathles09} & break-up & $0.006\pm0.0006$ & $17.5\pm1.8$ & $(387,478,579)$ & 600 & $250\pm85$ & $(1,3,6)$ & $0.4\pm0.3$ \\
			\citet{Collins15} & break-up & $0.9\pm0.09$ & $12\pm1.2$ & $(181,220,261)$ & 80 & $0.55\pm0.18$ & $(0.4,1.25,3)$ & $0.26\pm0.15$ \\
			\citet{Liu88} & break-up & $2\pm0.5$ & $18\pm1.8$ & $(188,250,313)$ & $^{(2)}$ & $(0.4,0.8,1.6)$ & $(1,3,6)$ & $0.4\pm0.3$ \\
			\citet{Sutherland16} & break-up & $0.08\pm0.008$ & $7.14\pm0.7$ & $80\pm16$ & 85 & $0.55\pm0.18$ & $(0.4,1.25,3)$ & $0.26\pm0.15$ \\
			\citet{Marchenko11} & break-up & $0.2\pm0.02$ & $7\pm0.7$ & $28\pm3$ & 1.75 & $0.5\pm0.17$ & $(0.4,1.25,3)$ & $0.25\pm0.05$ \\
			\citet{Marchenko12} & break-up & $0.31\pm0.03$ & $93\pm9.3$ & $1990\pm200$ & 47 & $0.94\pm0.31$ & $(0.2,0.77,1.6)$ & $0.339\pm0.034$ \\
			\citet{Marchenko12} & break-up & $0.14\pm0.014$ & $12.6\pm1.3$ & $217\pm32$ & 47 & $0.94\pm0.31$ & $(0.2,0.77,1.6)$ & $0.339\pm0.034$ \\
			\citet{Marchenko19} & no break-up & $0.03\pm0.003$ & $8\pm0.8$ & $(81,100,121)$ & 160 & $0.3\pm0.1$ & $(1.32,1.61,1.89)$ & $0.335\pm0.025$ \\ 
			\citet{Herman18} & no break-up & $0.02$ & $1.27$ & $2.52$ & $^{(2)}$ & $0.03$ & $0.009$ & $0.048$ \\
			\citet{Herman18} & break-up & $0.05$ & $1.27$ & $2.52$ & $^{(2)}$ & $0.03$ & $0.009$ & $0.048$ \\
			\citet{Herman18} & break-up & $0.07$ & $1.27$ & $2.52$ & $^{(2)}$ & $0.03$ & $0.009$ & $0.048$ \\
			\citet{Herman18} & break-up & $0.1$ & $1.27$ & $2.52$ & $^{(2)}$ & $0.03$ & $0.009$ & $0.048$ \\
			\citet{Herman18} & break-up & $0.1$ & $1.5$ & $3.51$ & $^{(2)}$ & $0.03$ & $0.009$ & $0.048$ \\
			\citet{Herman18} & no break-up & $0.01$ & $2$ & $6.17$ & $^{(2)}$ & $0.035$ & $0.016$ & $0.042$ \\
			\citet{Herman18} & no break-up & $0.01$ & $1.6$ & $3.99$ & $^{(2)}$ & $0.035$ & $0.016$ & $0.042$ \\
			\citet{Herman18} & no break-up & $0.01$ & $1.2$ & $2.25$ & $^{(2)}$ & $0.035$ & $0.016$ & $0.042$ \\
			\citet{Herman18} & no break-up & $0.02$ & $2$ & $6.17$ & $^{(2)}$ & $0.035$ & $0.016$ & $0.042$ \\
			\citet{Herman18} & no break-up & $0.03$ & $2$ & $6.17$ & $^{(2)}$ & $0.035$ & $0.016$ & $0.042$ \\
			\citet{Herman18} & no break-up & $0.04$ & $2$ & $6.17$ & $^{(2)}$ & $0.035$ & $0.016$ & $0.042$ \\
			\citet{Herman18} & no break-up & $0.05$ & $2$ & $6.17$ & $^{(2)}$ & $0.035$ & $0.016$ & $0.042$ \\
			\citet{Herman18} & no break-up$^{(1)}$ & $0.07$ & $2$ & $6.17$ & $^{(2)}$ & $0.035$ & $0.016$ & $0.042$ \\
			\citet{Herman18} & break-up & $0.09$ & $2$ & $6.17$ & $^{(2)}$ & $0.035$ & $0.016$ & $0.042$ \\
			\citet{Herman18} & break-up & $0.05$ & $1.6$ & $3.99$ & $^{(2)}$ & $0.035$ & $0.016$ & $0.042$ \\
			\citet{Herman18} & break-up & $0.07$ & $1.6$ & $3.99$ & $^{(2)}$ & $0.035$ & $0.016$ & $0.042$
		\end{tabular}
		\caption{Wave and ice properties used to calculate the break-up number $I_{br}$ for observed wave-ice events. Uncertainty in variables is taken into account through a triangular probability distribution, defined by the minimum, most probable and maximum value, respectively. Notes: $^{(1)}$while break-up was observed by the authors, \citet{Herman18} argue this was induced by reflecting waves rather than the incoming waves and, as such, this experiment is treated as a non-break-up event; $^{(2)}$water depth is not required to determine the characteristic wave length as this is given by the authors directly; $^{(3)}$Due to a typographical error in \citet{Asplin12} their estimated flexural strength $\sigma=40.7$ KPa is incorrect, the correct value should be 0.39 MPa.}
		\label{tab:Table1}
	\end{sidewaystable}
	
	\section{Results}
	\label{sec:Results}
	
	\subsection{Antarctic deployment}
	
	\begin{figure}
		\centerline{\includegraphics[]{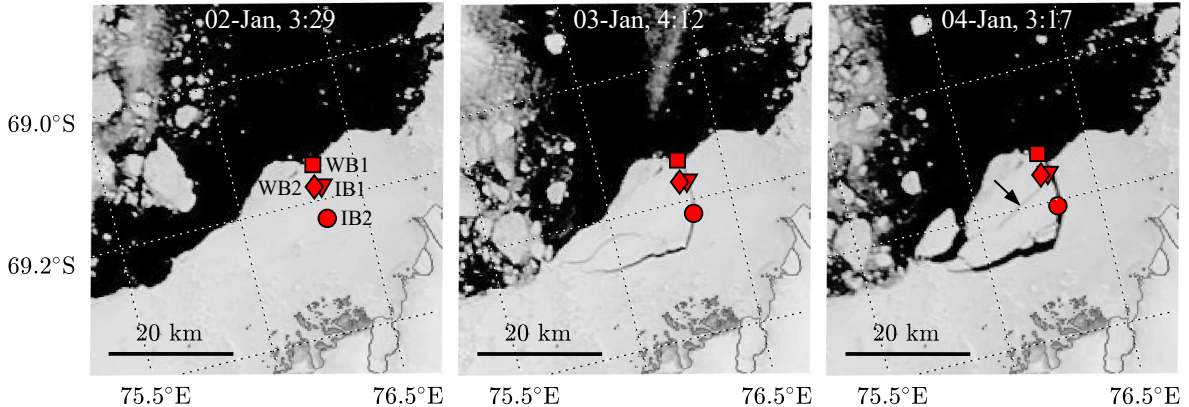}}
		\caption{MODIS imagery (https://worldview.earthdata.nasa.gov/) of the Antarctic deployment site on three consecutive cloud-free days during the initial sea ice break-up. Instruments are indicated by markers: WB1 (square); WB2 (diamond); IB1 (triangle); IB2 (circle). Note that the marker of IB1 is shifted here for visualization purposes, and that IB1 was originally deployed 40 m from WB2.}
		\label{fig:Breakup_initial}
	\end{figure}
	
	The first break-up event observed during the Antarctic campaign occurred about three weeks after instrument deployment. Based on satellite images, it can be observed that between  02 and 03-01-2020 a giant ice floe (approximately $20\times 10$ km in size) broke from the fast ice (see Fig. \ref{fig:Breakup_initial}). Based on the sudden change in geographical location of all four instruments (not shown here), this occurred around 01-01-2020 18:00. It also shows that all instruments are located on this giant ice floe which drifted at an average speed of approximately 0.03 m/s after the initial break-up. Note that on the satellite images of the 03 and 04-01-2020 multiple cracks can be observed (see arrow in Fig. \ref{fig:Breakup_initial}). Unfortunately, clouds in the days after prevent us from monitoring the ice conditions in the days that followed.
	
	\begin{figure}
		\centerline{\includegraphics[]{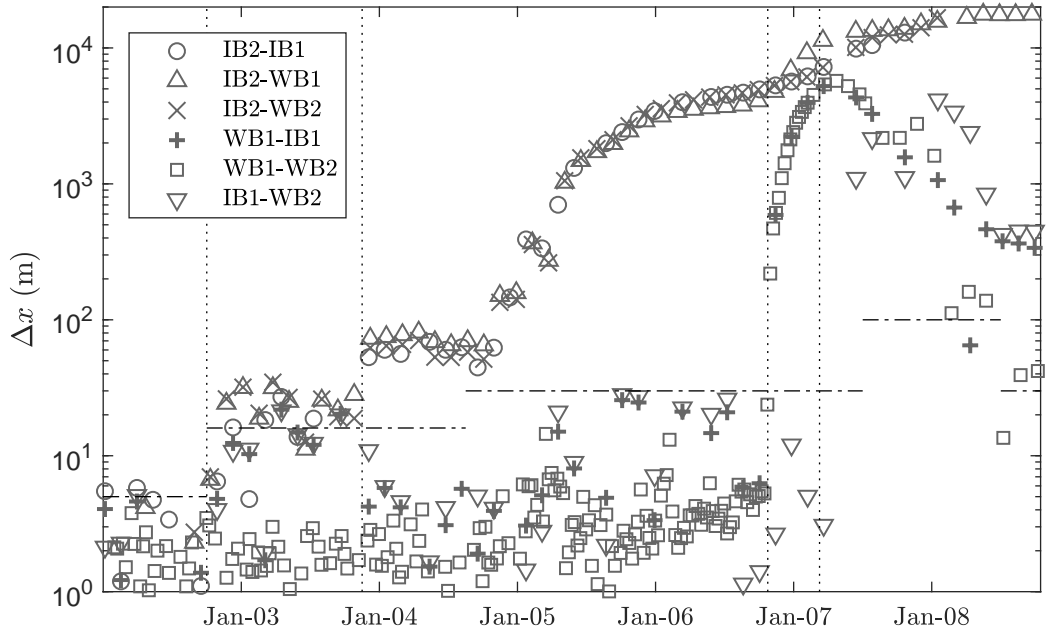}}
		\caption{Distance $\Delta x$ between instruments relative to their distance at the time of deployment, i.e., $\Delta x$ is taken initially equal to 0, and any change to $\Delta x$ indicates relative motion between the instruments. Vertical dashed lines indicate instances of sea ice break-up: (02-01-2020 18:00) all instruments start drifting due to break-up of large ice mass; (03-01-2020 21:00) IB2 separates from the large ice mass, see also Fig. \ref{fig:Breakup_initial}c; (06-01-2020 19:30) WB1 splits from WB2 and IB1; (07-01-2020 04:30) the ice floe holding WB2 and IB1 breaks due to waves generated by the storm depicted in Fig. \ref{fig:Weather}b. Horizontal dashed lines refer to the uncertainty level induced by GPS accuracy and interpolation error, where the latter increases with the drift speed of the instruments.}
		\label{fig:Breakup_Distance}
	\end{figure}
	
	As all instruments transmit their geographical location at regular intervals, albeit at different times, we can identify the occurrence of sea ice break-up and approximate the times at which these events occurred through the monitoring of sudden changes in the relative distance between buoy-pairs $\Delta x$ during the deployment (see Fig. \ref{fig:Breakup_Distance}). In all the following, the distance $\Delta x$ is relative to the distance at the time of deployment, i.e., initially $\Delta x$ is taken equal to 0, and any change in $\Delta x$ is due to relative motion of the instruments. However, for brevity, we will refer to this quantity as the `distance' between the instruments.
	
	As the geographical coordinates of the instruments are not transmitted at the same time and interval, we linearly interpolate the latitude and longitude coordinates to match between buoy-pairs. As the ice floe upon which the instruments rest drifts, interpolation of the geographical location introduces a maximum error of typical magnitude $\left|\delta \right|\approx (\Delta t^2/8)\hspace{1mm}\text{max}\left|\Delta x'' (t)\right|$, where $\Delta t$ is the data transmission interval. The estimated value of the error $\delta$ is indicated by the horizontal dashed lines in Fig. \ref{fig:Breakup_Distance}. Before the first sea ice break-up event, the approximate maximum error of $\Delta x$ is 5 m, a result of the accuracy of the GPS units when kept stationary during the initial three weeks of the deployment. From the instant at which the giant ice floe breaks from the ice cover and starts drifting (02-01-2020 18:00), the error increases to typically $\delta =16$ m. Note that the distance between all buoy-pairs remains constant just after the separation event of the giant ice floe, as all instruments remain on the one ice floe. Also note that the accuracy of the distance between the two wave buoys is considerably better than with other buoy-pairs, as the data transmission interval $\Delta t$ is considerably smaller for the wave buoys than for the ice buoys.
	
	After the giant ice floe separates from the ice cover and starts drifting, the next break-up event is thought to occur around the 03-01-2020 at 21:00, where the distance between IB2 and the other three buoys instantly increases to a distance of 60 --70 m (Fig. \ref{fig:Breakup_Distance}). This is in line with the satellite imagery (Fig. \ref{fig:Breakup_initial}), where on 03-01-2020 the crack does not seem to have propagated all the way eastward, whereas on the 04-01-2020 the crack seems to have split the giant ice floe completely (see the arrow, Fig. \ref{fig:Breakup_initial}). It is not until the 05-01-2020 that the distance between IB2 and the other instruments increases further. The third break-up event occurred around 06-01-2020 19:30, where the northernmost deployed instrument, WB1, splits from WB2 and IB1 (Fig. \ref{fig:Breakup_initial}). This is followed shortly after by a fourth break-up event occurring around 07-01-2020 4:30 where the distance between WB2 and IB1 increases to about a kilometer within just 3 hours.
	
	\begin{figure}
		\centerline{\includegraphics[]{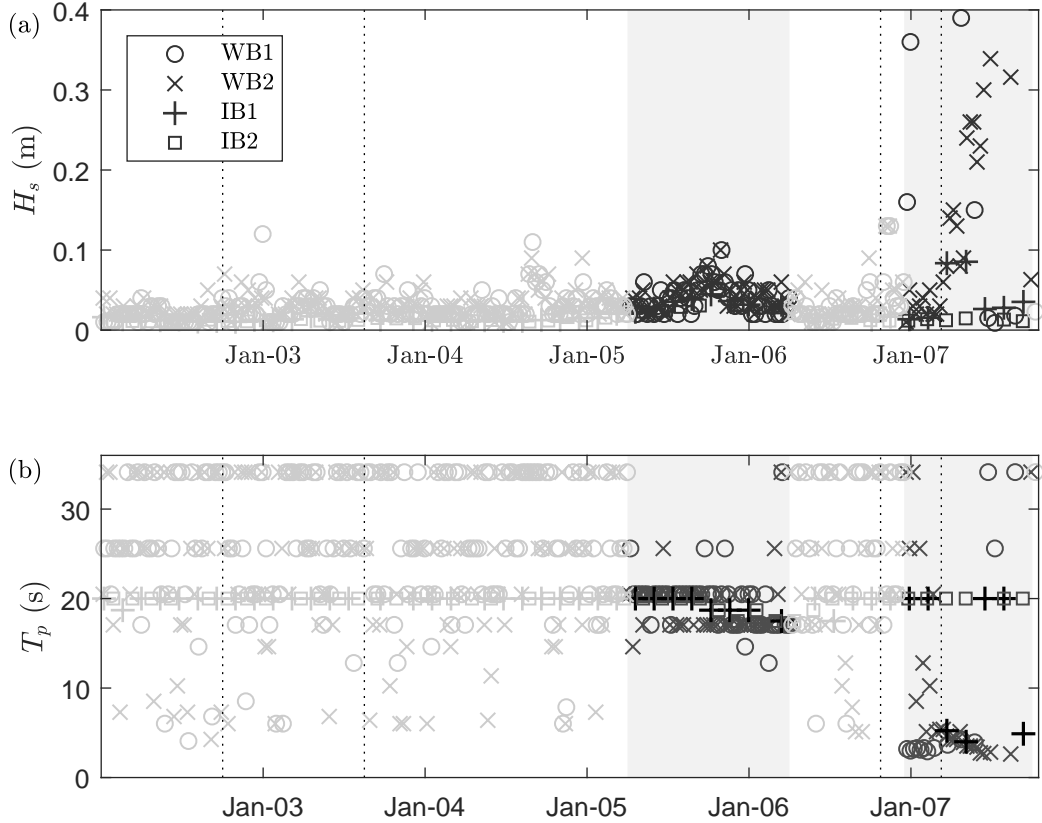}}
		\caption{(a) Significant wave height and (b) peak period measured by the four instruments during the break-up of the Antarctic fast sea ice cover. Based on consistency of the measured peak period between all instruments, two sections contain reliable wave measurements over noise thresholds, corresponding to a swell event with maximum ice motion obtained on 05-01-2020 18:00, and wind waves just after 07-01-2020. These sections correspond to the greyed areas and dark markers. Note that the vertical dashed lines indicate sea ice break-up events, extracted from Fig. \ref{fig:Breakup_Distance}.}
		\label{fig:Breakup_Waves}
	\end{figure}
	
	To determine whether these break-up events were caused by wave-induced flexural motion, they are compared against the wave motions recorded by the instruments. Fig. \ref{fig:Breakup_Waves} shows the significant wave height and peak wave period measured by the instruments over a duration of six days after the initial break away of the giant ice floe. Note that up to 05-01-2020, the instruments do no provide reliable wave information as recorded motions are below the noise threshold of the instruments. While this can be observed indirectly from the transmitted $H_s$ and $T_p$, for the ice buoys this is confirmed through observation of the wave energy spectra, showing a linear energy decay in log-scale from low to high frequencies, which corresponds to the noise threshold of the IMU \citep{Rabault20}. There are, however, two clear instances of coherent measurements of both the peak period and wave height, see shaded areas in Fig. \ref{fig:Breakup_Waves}.
	
	For the first break-up event on the 02-01-2020, no waves were measured above the noise level of the instruments and the cause of the break away of the giant ice floe remains speculative. ERA5 re-analysis data just north of the most northern sea ice edge indicates the presence of a 3 m swell a few hours preceding the break-up (generated by the storm depicted in Fig. \ref{fig:Weather}a), and, as such, swell might have been a potential cause of the break-up. However, as no significant ice motion events were recorded during this period of time by the instruments, it suggests that this swell event was largely dissipated by the vast sea ice band in front of the polynya. As there are no reliable wave measurements for the second and third break-up events either, we can only speculate about the cause of these events as well. As a few large cracks in the giant ice floe are already visible on the 03-01-2020 (Fig. \ref{fig:Breakup_initial}), therefore, it is most likely that the second break-up event was initiated at the same instant at which the giant ice floe broke from the fast ice cover. The third break-up event, however, is most likely induced by waves generated by the more energetic storm passing the deployment site during this part of the deployment (Fig. \ref{fig:Weather}b).
	
	\begin{figure}
		\centerline{\includegraphics[]{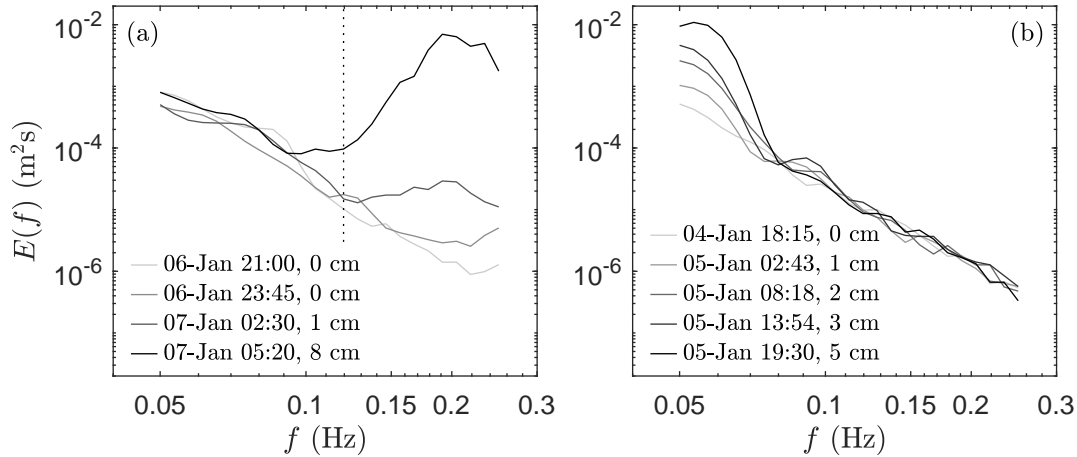}}
		\caption{Wave energy spectra measured by (a) IB1 during the fourth identified sea ice break-up event, and (b) IB2 during a swell event without sea ice break-up. The significant wave height of the spectra is provided in the legend, note, for (a) only the high frequency part (i.e. $f>0.13$) of the spectrum is considered, see dashed line. The spectra obtained on the 06-01-2020 21:00 (a) and 04-01-2020 18:15 (b) correspond to the noise level of the IMU. Both the measured wind waves in (a) and swell in (b) are well above noise levels. Note that in (a) energy in the high frequency part of the spectrum increases substantially after ice breakup, which is estimated to take place around 07-Jan 4:30.}
		\label{fig:Breakup_Spectrum}
	\end{figure}
	
	Unlike the first three break-up events, wave motions above noise thresholds were measured during the fourth sea ice break-up event. In particular, this break-up event coincides with the passage of the low pressure system and the presence of high wind speeds of about 10 -- 15 m/s, over and aligned with the main axis of the polynya region (this based on ERA5, see Fig. \ref{fig:Weather}b). With an area of approximately $100 \times 300$ km, the polynya provide sufficient fetch for the waves to develop. Around the time of break-up, a consistent peak wave period of around 5 seconds is measured by WB2 and, to lesser extent, by IB1. The wave energy spectra measured by WB2, however, shows that the wave energy in the high frequency range (around $f=0.2$) increases steadily with time (Fig. \ref{fig:Breakup_Spectrum}a). This explains for the sudden change in $T_p$ for IB1: the noise level at the lowest resolved frequency is larger than the measured wave energy in the high frequency range, so the wave amplitude of the relatively high-frequency waves has to reach a threshold before it is considered as the peak wave frequency $T_p$.
	
	The significant wave height of the high frequency waves (that is, when considering the wave energy for $f>0.12$) is only 0.01 m at 2:30 on 07-01-2020, and 0.08 m at 5:20. This suggests that the fourth break-up event, occurring around 4:30, was induced by waves with period of approximately 5 s, with an estimated wave height of around 0.04 m. It is noteworthy that the wave buoy WB1, which separated from WB2 and IB1 during the third break-up event, measured a significant wave height of up to 0.4 m at the time of the fourth break-up event, also with a period of approximately 4--5 s, indicating that the energetic wind waves were generated locally (since, if generated in the Southern Ocean, these waves would have dissipated rapidly in the sea ice band north of the polynya).
	
	Besides this wave-induced break-up event, a distinct swell event around the 05-01-2020 18:00 was measured by all four instruments (Fig. \ref{fig:Breakup_Waves}), though, it did not lead to sea ice break-up. From the spectra measured by the ice buoys it can be seen that the observed wave energy is comfortably above instrument noise level (Fig. \ref{fig:Breakup_Spectrum}b). The time frame of this swell event corresponds well to the passage of a storm moving north-east at this instant. This swell event will be used as a non-break-up event with a significant wave height of 0.05 m and period $T=17-20$ s (Fig. \ref{fig:Breakup_Waves} and \ref{fig:Breakup_Spectrum}b).
	
	\subsection{Arctic experiment}
	
	\begin{figure}
		\centerline{\includegraphics[]{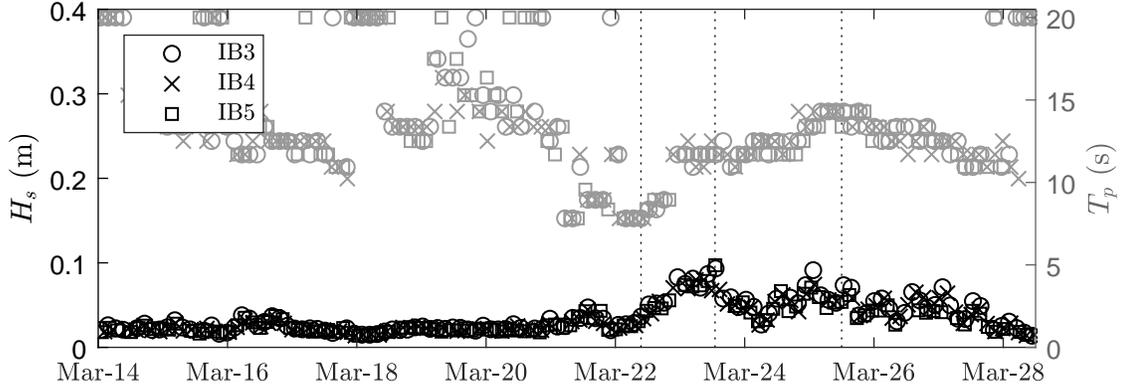}}
		\caption{Significant wave height and peak period measured by three ice buoys deployed on fast ice in Svalbard. The dashed lines identify three events with distinct peak wave period and peak significant wave height.}
		\label{fig:Arctic_Waves}
	\end{figure}
	
	During the Arctic field campaign, no sea ice break-up was observed and all instruments remained stationary during the deployment. The measurements of significant wave height and peak wave period are shown in Fig. \ref{fig:Arctic_Waves}. Three distinct wave events are considered as ice motion observations without sea ice break-up. The events have a peak period $T_p=7.8$, 11.7 and 14.3 s respectively, and corresponding wave heights are $H_s=4$, 0.10 and 0.07 m (see dashed lines in Fig. \ref{fig:Arctic_Waves}).
	
	\subsection{Ice break-up threshold}
	
	\begin{figure}
		\centerline{\includegraphics[]{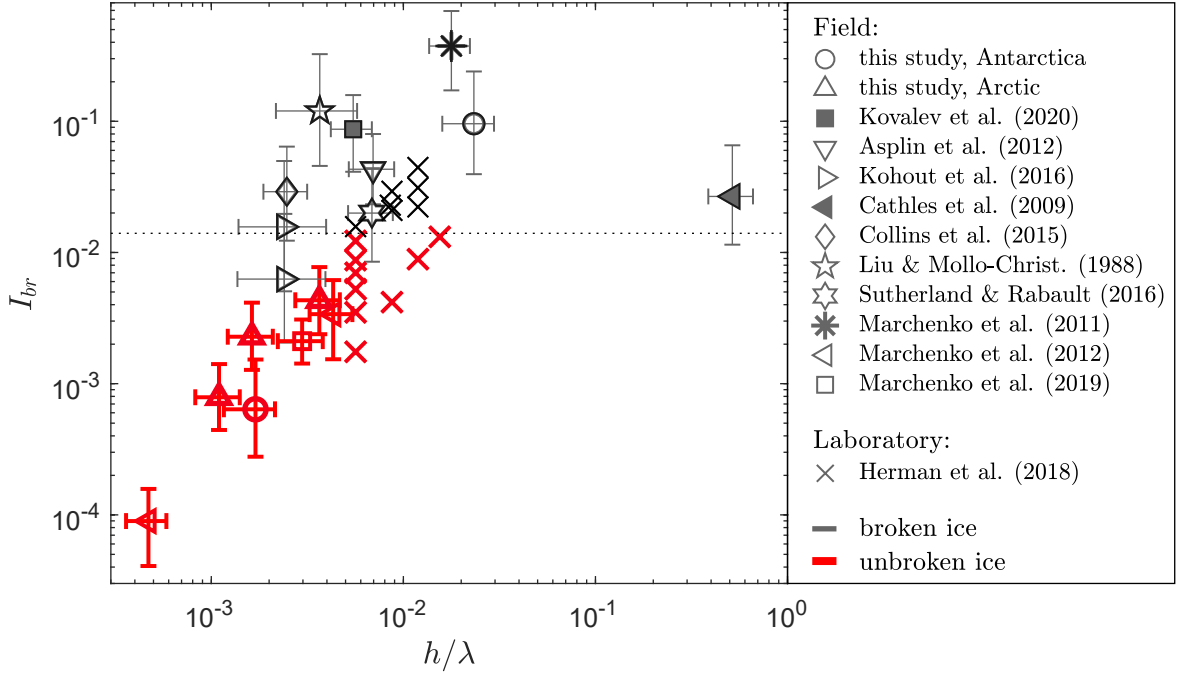}}
		\caption{Observations of $I_{br}$ against the relative ice thickness $h/\lambda$ for the complete data set. Events of wave-induced sea ice break-up are indicated with black markers, whereas observations where the flexural motion did not lead to break-up of the sea ice are shown with red markers. The observational threshold value $I_{br}\approx 0.014$, that separates the break-up from the non-break-up events, is indicated by the dashed line.}
		\label{fig:Breakup}
	\end{figure}
	
	Combining the break-up and non-break-up events obtained during the two field campaigns, and the set of existing published observations, the ice break-up parameter $I_{br}$ can be determined (see results in Fig. \ref{fig:Breakup}). In Fig. \ref{fig:Breakup}, we plotted $I_{br}$ against the relative ice thickness $h/\lambda$ to separate between ice breaking and non-breaking observations. Note that the red markers identify events where the ice remained intact under the wave motion. We reiterate that, similarly but contrapositive for the unbroken ice events, observations of sea ice break-up define a sufficient condition for wave-induced sea ice break-up, not the absolute threshold for the break-up parameter $I_{br}$. It is seen that broken and unbroken observations can be reasonably separated by a constant value of $I_{br}$. Therefore, based on the data presented in Fig. \ref{fig:Breakup}, we find the critical value of $I_{br}$ to be equal to:
	
	\begin{equation}
	I_{br}\approx 0.014.
	\label{eq:wave}
	\end{equation}
	
	While this threshold is most accurately boxed by the laboratory experiments of \citet{Herman18} (particularly as these constitute about half of the points in the data set), observations obtained in the field are well aligned with this threshold too. Note that, while one of the shipborne break-up observations of \citet{Kohout16} falls below this threshold, the large uncertainty of this particular visual observation covers both sides of the critical threshold. 
	
	\section{Discussion}
	\label{sec:Discussion}
	
	In the present work, we have collected experimental observations, both from the laboratory and the from field, displaying both wave-induced sea ice break-up events and wave-induced ice motion events without break-up. Thereafter, we have used these data to estimate the critical threshold value for the wave-induced sea ice break-up parameter $I_{br}$. We find that observations consistently point to a constant value of $I_{br}$, which we estimate to be $I_{br}\approx 0.014$ (see Fig. \ref{fig:Breakup}). Note, however, as we can only measure statistical wave properties in an incoherent wave field, and thus $I_{br}$ is a probabilistic metric rather than a deterministic, the threshold observed in this study therefore suggests that above $I_{br}=0.014$ the ice is very likely (but not necessarily) going to break. Though the data set is still rather limited, it is promising that both field and laboratory observations are well aligned with this critical value. In particular, laboratory-grown ice is known to have distinctly different material properties \citep[e.g.,][]{Herman18,Squire20} where, for instance, the ice in the laboratory data used here has a critical strain one to two orders of magnitude larger than that of sea ice in the field. Interestingly, the swell-induced crack propagation of the Ross Ice Shelf, as implied by \citet{Cathles09} and \citet{Massom18}, seems to fit well within the overall dataset, indicating that it might be possible to extrapolate the wave-induced sea ice break-up criterion to much thicker ice covers as well.
	
	While we observe that the critical value determined in this study is three to four times smaller than that of a monochromatic wave \citep[e.g.][]{Williams13a}, this value is remarkably similar to that proposed by \citet{Boutin18}, who argued, based on statistical considerations, that a factor of 3.6 should be used to take into account the random nature of the wave field and the resulting stochastic distribution in individual wave amplitudes. However, as the ice in the laboratory experiments of \citet{Herman18} were exposed to monochromatic waves, rather than a random wave field, it remains uncertain whether this factor is indeed a statistical correction, a compensation for the simplification of the sea ice material properties (that is, by ignoring fatigue and the presence of sea ice heterogeneities, the critical flexural strength of the ice is effectively lower than those values used here), or, more likely, a combination of both. Either way, our experimental results are in support of the current approaches developed to model the break-up of a solid ice cover under wave forcing in coupled numerical models, albeit further study is required to understand the finer details of the physics behind wave-induced sea ice break-up.
	
	Although the wave-induced sea ice break-up parameter $I_{br}$ seems to be physically sound, the scaling of $I_{br}\propto h$ is problematic when the ice material properties $Y$ and $\sigma$ remain virtually unchanged when thin ice is considered, that is, an infinitely thin ice sheet becomes numerically unbreakable \citep[as noted in:][]{WWIII}. However, for small ice thicknesses, other physical processes may be naturally dominant, such as compressive or tensile failure of the ice through wind and ocean current shear forces. Indeed, the relative effect of such forcing scales inversely to the ice thickness \citep[e.g.,][]{Mellor86}, contrary to what is obtained with the present expression for $I_{br}$. This highlights that waves and sea ice are part of a complex coupled system at the interface of the atmosphere and ocean, and that many different physical phenomena influence sea ice dynamics. Waves can, however, still play a critical role in the break-up of thin ice. For instance, thick ice attenuates wave energy more strongly than thin ice \citep[e.g.,][]{Doble15,Meylan18,Liu20}, therefore, thin ice is generally exposed to more wave energy, including shorter wavelengths. Moreover, there are still significant uncertainties in the actual mechanical properties of very thin ice relative to thicker ice. Fast grown thin ice (for instance, in the case of very cold air temperature) has a lower flexural strength compared to slow grown ice \citep{Bond97}, which the literature claims to be caused by its higher bulk salinity \citep{Perovich94}. Moreover, \cite{Kovacs96} finds that the salinity of young ice decreases with increasing ice thickness, implying that thin ice might be consistently weaker than thicker ice following Eqn.\hspace{1mm}(\ref{eq:sigma_Karulina}) and (\ref{eq:sigma_Timco}). As ice properties can vary significantly in time, more studies are required to accurately measure and define the mechanical properties of sea ice in terms of more readily available air-sea-ice properties, and the role of ice inhomogeneities caused by bubbles and brine pockets, ice ridges, pools, and ice thickness variability, needs to be further investigated.
	
	Field observations of waves, ice motion, ice material properties, and sea ice break-up identification, bring unavoidable uncertainties, resulting in a significant uncertainty for $I_{br}$. Particularly the mechanical properties of sea ice are uncertain due to the validity of the experimental methods used \citep[e.g., see][]{Timco10,karulin2019features}, fatigue \citep[e.g.,][]{Langhorne98}, spatial heterogeneity at various scales and even questions regarding the scaling effects of the ice flexural strength \citep{Aly19}. Identifying the instant at which the ice breaks creates an additional uncertainty. The method which consists in identifying the instant of sea ice break-up through the spatial divergence of instrumentation, as applied in this study, is not foolproof by itself. In fact, if the ice floes do not drift apart after break-up, the relative distance between instruments will not change. As the sea ice in our field experiments was drifting during break-up, it is expected that the resulting ice floes after break-up will attain a different drift speed. In the case of the Antarctic field campaign, the instruments drifted at a speed of 0.03 -- 0.20 m/s and, even if the differential drift between floes immediately after break-up is only a fraction of this drift speed, this will be noticed from the position of the instruments within hours of the time of break-up, at most.
	
	Therefore, a dedicated field experiment, with the aim to closely monitor both the mechanical properties of the ice and the exact instant at which the ice breaks, is highly desirable and is expected to provide further clarification over the accuracy of the observed threshold for $I_{br}$ reported here. Until then, many more observations of wave-induced ice motion leading up to ice break-up are necessary to further substantiate the wave-induced break-up parameter and its critical threshold. Evidently, development of low-cost and open source instrumentation is critical in obtaining a large dataset of break-up observations, as it promotes the deployment of ice buoys in larger quantities and, therefore, allows to dramatically increase the overall volume of data reporting the interactions between sea ice, waves, atmosphere, and the ocean.
	
	\section{Conclusions}
	\label{sec:Conclusions}
	
	We presented observations of wave-induced ice motion and sea ice break-up events from two field experiments, one in the Antarctic and the other in the Arctic. Using the relative displacement between the instruments deployed, four sea ice break-up events were registered in the Antarctic field experiment, although only one could, with reasonably certainty, be linked to waves. While no sea ice break-up events were observed in the Arctic field experiments, it provided three wave events without sea ice break-up. We used these observations, supplemented with existing data taken from a wide body of the literature, to reach an estimate for the critical threshold of the wave-induced sea ice break-up parameter $I_{br}=ahY/\sigma\lambda^2$, where $a$ is the wave amplitude, $h$ is the ice thickness, $Y$ is the Young's Modulus, $\sigma$ is the ice flexural strength, and $\lambda$ is the wave length. We find that a value $I_{br} = 0.014$ separates well observations of wave-induced break-up and non break-up events. Observations include laboratory measurements, as well as suspected cracking of the Antarctic ice shelf. The physical relevance of $I_{br}$ is substantiated by the diversity of cases present in the data, from laboratory to the field, the Antarctic to the Arctic, and thin ice to very thick ice. However, significantly more observations of sea ice break-up are necessary and, perhaps, more sophisticated measurement techniques need to be developed, in order to identify the exact instant at which break up occurs, and the wave conditions responsible for the observed sea ice break-up.
	
	\section*{Acknowledgements}
	
	We acknowledge the use of imagery from the NASA Worldview application (https://worldview.earthdata.nasa.gov/), part of the NASA Earth Observing System Data and Information System (EOSDIS). Authors would like to thank the crew of AARI for their assistance during the deployment of the instruments used in the Antarctic. We thank Prof. Atle Jensen and Ing. Olav Gundersen for their support in assembling IB1 and IB2 (IB3, IB4, and IB5 were assembled at Melbourne University following the same design). Data collection in Gr\o{}nfjorden, Svalbard was conducted within the expedition `Spitsbergen-2020' organised by Russian Scientific Arctic Expedition on Spitsbergen Archipelago (RAE-S), AARI. JJV and AVB acknowledge support from the Joyce Lambert Antarctic Research Fund (Grant 604086); JJV, AVB, and PH were supported by the Australian Antarctic Program under Project 4593 plus PH under 4506; JJV, JR, KF, AM, and AVB acknowledge the support of the Research Council of Norway through the SFI SIB project. JR was supported in the context of DOFI project (Univ. of Oslo, Grant Number 280625).
	
	
	\bibliography{Library_JJV}

@article{Spreen08,
	title={Sea ice remote sensing using AMSR-E 89-GHz channels},
	author={Spreen, Gunnar and Kaleschke, Lars and Heygster, Georg},
	journal={Journal of Geophysical Research: Oceans},
	doi={10.1029/2005JC003384},
	volume={113},
	number={C2},
	year={2008},
	publisher={Wiley Online Library}
}

@article{Kovalev20,
	title={Crack formation and breakout of shore fast sea ice in Mordvinova Bay, south-east Sakhalin Island},
	author={Kovalev, Dmitry P and Kovalev, Peter D and Squirecor, Vernon A},
	journal={Cold Regions Science and Technology},
	pages={103082},
	year={2020},
	publisher={Elsevier}
}

@article{Asplin12,
	title={Fracture of summer perennial sea ice by ocean swell as a result of Arctic storms},
	author={Asplin, Matthew G and Galley, Ryan and Barber, David G and Prinsenberg, Simon},
	journal={Journal of Geophysical Research: Oceans},
	volume={117},
	number={C6},
	year={2012},
	publisher={Wiley Online Library}
}

@article{Kohout16,
	title={In situ observations of wave-induced sea ice breakup},
	author={Kohout, AL and Williams, MJM and Toyota, Takenobu and Lieser, J and Hutchings, J},
	journal={Deep Sea Research Part II: Topical Studies in Oceanography},
	volume={131},
	pages={22--27},
	year={2016},
	publisher={Elsevier}
}

@article{Cathles09,
	title={Seismic observations of sea swell on the floating Ross Ice Shelf, Antarctica},
	author={Cathles, LM and Okal, Emile A and MacAyeal, Douglas R},
	journal={Journal of Geophysical Research: Earth Surface},
	volume={114},
	number={F2},
	year={2009},
	publisher={Wiley Online Library}
}

@article{Collins15,
	title={In situ measurements of an energetic wave event in the Arctic marginal ice zone},
	author={Collins, Clarence O and Rogers, W Erick and Marchenko, Aleksey and Babanin, Alexander V},
	journal={Geophysical Research Letters},
	volume={42},
	number={6},
	pages={1863--1870},
	year={2015},
	publisher={Wiley Online Library}
}

@article{Liu88,
	title={Wave propagation in a solid ice pack},
	author={Liu, Antony K and Mollo-Christensen, Erik},
	journal={Journal of physical oceanography},
	volume={18},
	number={11},
	pages={1702--1712},
	year={1988}
}

@article{Bromirski10,
	title={Transoceanic infragravity waves impacting Antarctic ice shelves},
	author={Bromirski, Peter D and Sergienko, Olga V and MacAyeal, Douglas R},
	journal={Geophysical Research Letters},
	volume={37},
	number={2},
	year={2010},
	publisher={Wiley Online Library}
}

@article{Massom18,
	title={Antarctic ice shelf disintegration triggered by sea ice loss and ocean swell},
	author={Massom, Robert A and Scambos, Theodore A and Bennetts, Luke G and Reid, Phillip and Squire, Vernon A and Stammerjohn, Sharon E},
	journal={Nature},
	volume={558},
	number={7710},
	pages={383--389},
	year={2018},
	publisher={Nature Publishing Group}
}

@article{Sutherland16,
	title={Observations of wave dispersion and attenuation in landfast ice},
	author={Sutherland, Graig and Rabault, Jean},
	journal={Journal of Geophysical Research: Oceans},
	volume={121},
	number={3},
	pages={1984--1997},
	year={2016},
	publisher={Wiley Online Library}
}

@inproceedings{Marchenko11,
	title={Field studies of sea water and ice properties in Svalbard fjords},
	author={Marchenko, A and Shestov, A and Karulin, E and Morozov, E and Karulina, M and Bogorodsky, P and Muzylev, S and Onishchenko, D and Makshtas, A},
	booktitle={Proceedings of the International Conference on Port and Ocean Engineering Under Arctic Conditions},
	number={POAC11-148},
	year={2011}
}

@article{Herman18,
	title={Floe-size distributions in laboratory ice broken by waves.},
	author={Herman, Agnieszka and Evers, Karl-Ulrich and Reimer, Nils},
	journal={Cryosphere},
	volume={12},
	number={2},
	year={2018}
}

@article{Williams13a,
	title={Wave--ice interactions in the marginal ice zone. Part 1: Theoretical foundations},
	author={Williams, Timothy D and Bennetts, Luke G and Squire, Vernon A and Dumont, Dany and Bertino, Laurent},
	journal={Ocean Modelling},
	volume={71},
	pages={81--91},
	year={2013},
	publisher={Elsevier}
}

@article{Williams13b,
	title={Wave--ice interactions in the marginal ice zone. Part 2: Numerical implementation and sensitivity studies along 1D transects of the ocean surface},
	author={Williams, Timothy D and Bennetts, Luke G and Squire, Vernon A and Dumont, Dany and Bertino, Laurent},
	journal={Ocean Modelling},
	volume={71},
	pages={92--101},
	year={2013},
	publisher={Elsevier}
}

@article{Dumont11,
	title={A wave-based model for the marginal ice zone including a floe breaking parameterization},
	author={Dumont, D and Kohout, A and Bertino, L},
	journal={Journal of Geophysical Research: Oceans},
	volume={116},
	number={C4},
	year={2011},
	publisher={Wiley Online Library}
}

@article{Voermans19,
	title={Wave attenuation by sea ice turbulence},
	author={Voermans, JJ and Babanin, AV and Thomson, J and Smith, MM and Shen, HH},
	journal={Geophysical Research Letters},
	volume={46},
	number={12},
	pages={6796--6803},
	year={2019},
	publisher={Wiley Online Library}
}

@article{Timco94,
	title={Flexural strength equation for sea ice},
	author={G.W. Timco and S. O'{B}rien},
	journal={Cold Regions Science and Technology},
	volume={22},
	number={3},
	pages={285--298},
	year={1994},
	publisher={Elsevier}
}

@article{Frankenstein67,
	title={Equations for determining the brine volume of sea ice from- 0.5° to- 22.9° C.},
	author={Frankenstein, Guenther and Garner, Robert},
	journal={Journal of Glaciology},
	volume={6},
	number={48},
	pages={943--944},
	year={1967},
	publisher={Cambridge University Press}
}

@article{Kohout14,
	title={Storm-induced sea-ice breakup and the implications for ice extent},
	author={Kohout, AL and Williams, MJM and Dean, SM and Meylan, MH},
	journal={Nature},
	volume={509},
	number={7502},
	pages={604--607},
	year={2014},
	publisher={Nature Publishing Group}
}

@article{Raghukumar19,
	title={Performance characteristics of “Spotter,” a newly developed real-time wave measurement buoy},
	author={Raghukumar, Kaustubha and Chang, Grace and Spada, Frank and Jones, Craig and Janssen, Tim and Gans, Andrew},
	journal={Journal of Atmospheric and Oceanic Technology},
	volume={36},
	number={6},
	pages={1127--1141},
	year={2019}
}

@article{Rabault20,
	title={An open source, versatile, affordable waves in ice instrument for scientific measurements in the Polar Regions},
	author={Rabault, Jean and Sutherland, Graig and Gundersen, Olav and Jensen, Atle and Marchenko, Aleksey and Breivik, {\O}yvind},
	journal={Cold Regions Science and Technology},
	volume={170},
	pages={102955},
	year={2020},
	publisher={Elsevier}
}

@article{Rabault16,
	title={Measurements of waves in landfast ice using inertial motion units},
	author={Rabault, Jean and Sutherland, Graig and Ward, Brian and Christensen, Kai H and Halsne, Trygve and Jensen, Atle},
	journal={IEEE Transactions on Geoscience and Remote Sensing},
	volume={54},
	number={11},
	pages={6399--6408},
	year={2016},
	publisher={IEEE}
}

@article{Rabault19,
	title={Experiments on wave propagation in grease ice: combined wave gauges and particle image velocimetry measurements},
	author={Rabault, Jean and Sutherland, Graig and Jensen, Atle and Christensen, Kai H and Marchenko, Aleksey},
	journal={Journal of Fluid Mechanics},
	volume={864},
	pages={876--898},
	year={2019},
	publisher={Cambridge University Press}
}

@article{Liu20,
	title={Spectral modelling of ice-induced wave decay},
	author={Liu, Qingxiang and Rogers, W Erick and Babanin, Alexander and Li, Jiangkai and Guan, Changlong},
	journal={Journal of Physical Oceanography},
	volume = {50},
	number = {6},
	pages = {1583-1604},
	year = {2020}
}

@article{Herman19,
	title={Wave energy attenuation in fields of colliding ice floes--Part 2: A laboratory case study},
	author={Herman, Agnieszka and Cheng, Sukun and Shen, Hayley H},
	journal={The Cryosphere},
	volume={13},
	number={11},
	pages={2901--2914},
	year={2019},
	publisher={Copernicus GmbH}
}

@article{Wang10,
	title={Gravity waves propagating into an ice-covered ocean: A viscoelastic model},
	author={Wang, Ruixue and Shen, Hayley H},
	journal={Journal of Geophysical Research: Oceans},
	volume={115},
	number={C6},
	year={2010},
	publisher={Wiley Online Library}
}

@article{Sutherland19_twolayer,
	title={A two layer model for wave dissipation in sea ice},
	author={Sutherland, Graig and Rabault, Jean and Christensen, Kai H and Jensen, Atle},
	journal={Applied Ocean Research},
	volume={88},
	pages={111--118},
	year={2019},
	publisher={Elsevier}
}

@article{Thomson18,
	title={Overview of the arctic sea state and boundary layer physics program},
	author={Thomson, Jim and Ackley, Stephen and Girard-Ardhuin, Fanny and Ardhuin, Fabrice and Babanin, Alex and Boutin, Guillaume and Brozena, John and Cheng, Sukun and Collins, Clarence and Doble, Martin and others},
	journal={Journal of Geophysical Research: Oceans},
	volume={123},
	number={12},
	pages={8674--8687},
	year={2018},
	publisher={Wiley Online Library}
}

@article{Collins18,
	title={Observations of surface wave dispersion in the marginal ice zone},
	author={Collins, Clarence and Doble, Martin and Lund, Bj{\"o}rn and Smith, Madison},
	journal={Journal of Geophysical Research: Oceans},
	volume={123},
	number={5},
	pages={3336--3354},
	year={2018},
	publisher={Wiley Online Library}
}

@article{Ardhuin18,
	title={Wave attenuation through an arctic marginal ice zone on 12 October 2015: 2. Numerical modeling of waves and associated ice breakup},
	author={Ardhuin, Fabrice and Boutin, Guillaume and Stopa, Justin and Girard-Ardhuin, Fanny and Melsheimer, Christian and Thomson, Jim and Kohout, Alison and Doble, Martin and Wadhams, Peter},
	journal={Journal of Geophysical Research: Oceans},
	volume={123},
	number={8},
	pages={5652--5668},
	year={2018},
	publisher={Wiley Online Library}
}

@article{Boutin20,
	title={Towards a coupled model to investigate wave--sea ice interactions in the Arctic marginal ice zone},
	author={Boutin, Guillaume and Lique, Camille and Ardhuin, Fabrice and Talandier, Claude and Accensi, Micka{\"e}l and Girard-Ardhuin, Fanny},
	journal={The Cryosphere},
	volume={14},
	number={2},
	pages={709--735},
	year={2020}
}

@article{Squire20,
	title={Ocean wave interactions with sea ice: a reappraisal},
	author={Squire, Vernon A},
	journal={Annual Review of Fluid Mechanics},
	volume={52},
	year={2020},
	publisher={Annual Reviews}
}

@article{Timco10,
	title={A review of the engineering properties of sea ice},
	author={Timco, GW and Weeks, WF},
	journal={Cold regions science and technology},
	volume={60},
	number={2},
	pages={107--129},
	year={2010},
	publisher={Elsevier}
}

@article{Stopa18,
	title={Strong and highly variable push of ocean waves on Southern Ocean sea ice},
	author={Stopa, Justin E and Sutherland, Peter and Ardhuin, Fabrice},
	journal={Proceedings of the National Academy of Sciences},
	volume={115},
	number={23},
	pages={5861--5865},
	year={2018},
	publisher={National Acad Sciences}
}

@article{Thomas19,
	title={Effect of wave-induced mixing on Antarctic sea ice in a high-resolution ocean model},
	author={Thomas, Steven and Babanin, Alexander V and Walsh, Kevin JE and Stoney, Lachlan and Heil, Petra},
	journal={Ocean Dynamics},
	volume={69},
	number={6},
	pages={737--746},
	year={2019},
	publisher={Springer}
}

@article{Langhorne98,
	title={Break-up of sea ice by ocean waves},
	author={Langhorne, Patricia J and Squire, Vernon A and Fox, Colin and Haskell, Timothy G},
	journal={Annals of Glaciology},
	volume={27},
	pages={438--442},
	year={1998},
	publisher={Cambridge University Press}
}

@article{Boutin18,
	title={Floe Size Effect on Wave-Ice Interactions: Possible Effects, Implementation in Wave Model, and Evaluation},
	author={Boutin, Guillaume and Ardhuin, Fabrice and Dumont, Dany and S{\'e}vigny, Caroline and Girard-Ardhuin, Fanny and Accensi, Mickael},
	journal={Journal of Geophysical Research: Oceans},
	volume={123},
	number={7},
	pages={4779--4805},
	year={2018},
	publisher={Wiley Online Library}
}

@techreport{WWIII,
	title={User manual and system documentation of {WAVEWATCH} {III} version 6.07. {T}ech. {N}ote 333},
	author={The{ }WAVEWATCH{ }{III}{ }Development{ }Group},
	institution={NOAA/NWS/NCEP/MMAB},
	year={2019},
	address={College Park, MD, USA}
}

@article{Meylan18,
	title={Dispersion relations, power laws, and energy loss for waves in the marginal ice zone},
	author={Meylan, Michael H and Bennetts, Luke G and Mosig, JEM and Rogers, WE and Doble, MJ and Peter, Malte A},
	journal={Journal of Geophysical Research: Oceans},
	volume={123},
	number={5},
	pages={3322--3335},
	year={2018},
	publisher={Wiley Online Library}
}

@article{Doble15,
	title={Relating wave attenuation to pancake ice thickness, using field measurements and model results},
	author={Doble, Martin J and De Carolis, Giacomo and Meylan, Michael H and Bidlot, Jean-Raymond and Wadhams, Peter},
	journal={Geophysical Research Letters},
	volume={42},
	number={11},
	pages={4473--4481},
	year={2015},
	publisher={Wiley Online Library}
}

@article{Bond97,
	title={Fatigue behavior of cantilever beams of saline ice},
	author={Bond, Paul E and Langhorne, Patricia J},
	journal={Journal of cold regions engineering},
	volume={11},
	number={2},
	pages={99--112},
	year={1997},
	publisher={American Society of Civil Engineers}
}

@article{Perovich94,
	title={Surface characteristics of lead ice},
	author={Perovich, Donald K and Richter-Menge, Jacqueline A},
	journal={Journal of Geophysical Research: Oceans},
	volume={99},
	number={C8},
	pages={16341--16350},
	year={1994},
	publisher={Wiley Online Library}
}

@techreport{Kovacs96,
	title={Sea ice. Part 1. Bulk salinity versus ice floe thickness},
	author={Kovacs, Austin},
	year={1996},
	institution={COLD REGIONS RESEARCH AND ENGINEERING LAB HANOVER NH}
}

@article{Aly19,
	title={Scale Effect in Ice Flexural Strength},
	author={Aly, Mohamed and Taylor, Rocky and Bailey Dudley, Eleanor and Turnbull, Ian},
	journal={Journal of Offshore Mechanics and Arctic Engineering},
	volume={141},
	number={5},
	year={2019},
	publisher={American Society of Mechanical Engineers Digital Collection}
}

@article{Steele92,
	title={Sea ice melting and floe geometry in a simple ice-ocean model},
	author={Steele, Michael},
	journal={Journal of Geophysical Research: Oceans},
	volume={97},
	number={C11},
	pages={17729--17738},
	year={1992},
	publisher={Wiley Online Library}
}

@inproceedings{Marchenko17,
	title={Three physical mechanisms of wave energy dissipation in solid ice},
	author={Marchenko, ALEKSEY and Cole, DAVID},
	booktitle={Proceedings of the 24th International Conference on Port and Ocean Engineering under Arctic Conditions, Busan, Korea},
	year={2017}
}

@article{Karulina19,
	title={Full-scale flexural strength of sea ice and freshwater ice in Spitsbergen Fjords and North-West Barents Sea},
	author={Karulina, M and Marchenko, A and Karulin, E and Sodhi, D and Sakharov, A and Chistyakov, P},
	journal={Applied Ocean Research},
	volume={90},
	pages={101853},
	year={2019},
	publisher={Elsevier}
}

@incollection{Mellor86,
	title={Mechanical behavior of sea ice},
	author={Mellor, Malcolm},
	booktitle={The geophysics of sea ice},
	pages={165--281},
	year={1986},
	publisher={Springer}
}

@article{Crocker89,
	title={Breakup of Antarctic fast ice},
	author={Crocker, GB and Wadhams, P},
	journal={Cold regions science and technology},
	volume={17},
	number={1},
	pages={61--76},
	year={1989},
	publisher={Elsevier}
}

@article{Marchenko19,
	title={Wave-ice interaction in the {N}orth-{W}est {B}arents {S}ea},
	author={Marchenko, Aleksey and Wadhams, Peter and Collins, C and Rabault, Jean and Chumakov, Mikhail},
	journal={Applied Ocean Research},
	volume={90},
	pages={101861},
	year={2019},
	publisher={Elsevier}
}

@techreport{Vaudrey77,
	title={Ice Engineering-Study of Related Properties of Floating Sea-Ice Sheets and Summary of Elastic and Viscoelastic Analyses},
	author={Vaudrey, K.D.},
	year={1977},
	institution={Civil Engineering Lab (Navy), Port Hueneme, CA}
}

@article{vaughan2007scattering,
	title={Scattering of ice coupled waves by a sea-ice sheet with random thickness},
	author={Vaughan, GL and Squire, VA},
	journal={Waves in Random and Complex Media},
	volume={17},
	number={3},
	pages={357--380},
	year={2007},
	publisher={Taylor \& Francis}
}

@article{horvat2016interaction,
	title={Interaction of sea ice floe size, ocean eddies, and sea ice melting},
	author={Horvat, Christopher and Tziperman, Eli and Campin, Jean-Michel},
	journal={Geophysical Research Letters},
	volume={43},
	number={15},
	pages={8083--8090},
	year={2016},
	publisher={Wiley Online Library}
}

@article{hwang2017winter,
	title={Winter-to-summer transition of Arctic sea ice breakup and floe size distribution in the Beaufort Sea},
	author={Hwang, Byongjun and Wilkinson, Jeremy and Maksym, Edward and Graber, Hans C and Schweiger, Axel and Horvat, Christopher and Perovich, Donald K and Arntsen, Alexandra E and Stanton, Timothy P and Ren, Jinchang and others},
	journal={Elementa Science of the Anthropocene},
	volume={5},
	year={2017},
	publisher={University of California Press}
}

@article{meylan2018three,
	title={Three-dimensional time-domain scattering of waves in the marginal ice zone},
	author={Meylan, MH and Bennetts, LG},
	journal={Philosophical Transactions of the Royal Society A: Mathematical, Physical and Engineering Sciences},
	volume={376},
	number={2129},
	pages={20170334},
	year={2018},
	publisher={The Royal Society Publishing}
}

@article{Murdza20,
	title={Strengthening of columnar-grained freshwater ice through cyclic flexural loading},
	author={Murdza, Andrii and Schulson, Erland M and Renshaw, Carl E},
	journal={Journal of Glaciology},
	pages={1--11},
	publisher={Cambridge University Press},
	year={2020}
}

@inproceedings{karulin2019features,
	title={Features of determining the ice flexural strength and the elastic modulus based on floating cantilever beam tests},
	author={Karulin, Evgeny B and Marchenko, Aleksey V and Sakharov, Aleksandr N and Karulina, Marina M and Chistyakov, Peter V and Onishchenko, Dmitry A},
	booktitle={Proceedings of the 25th International Conference on Port and Ocean Engineering under Arctic Conditions, June},
	year={2019}
}

@article{Marchenko12,
	title={A tsunami wave recorded near a glacier front},
	author={Marchenko, A. and Morozov, E.G. and Muzylev, S.V.},
	year={2012},
	journal={Natural Hazards and Earth System Sciences},
	volume={20},
	number={2},
	pages={415–-419}
}

@article{Marchenko13,
	title={Measurements of sea-ice flexural stiffness by pressure characteristics of flexural-gravity waves},
	author={Marchenko, Aleksey and Morozov, Eugene and Muzylev, Sergey},
	journal={Annals of Glaciology},
	volume={54},
	number={64},
	pages={51--60},
	year={2013},
	publisher={Cambridge University Press}
}

\end{document}